\documentclass[sn-mathphys]{sn-jnl}

\jyear{2022}%

\theoremstyle{thmstyleone}%
\newtheorem{theorem}{Theorem}
\newtheorem{proposition}[theorem]{Proposition}%

\theoremstyle{thmstyletwo}%

\theoremstyle{thmstylethree}%

\begin{document}

\title[Evaluating the Convergence of Tabu Enhanced Hybrid Quantum Optimization]{Evaluating the Convergence of Tabu Enhanced Hybrid Quantum Optimization}


\author*[1,2,3]{\fnm{Enrico} \sur{Blanzieri}}\email{enrico.blanzieri@unitn.it}

\author[1,2]{\fnm{Davide} \sur{Pastorello}}\email{d.pastorello@unitn.it}

\author[3]{\fnm{Valter} \sur{Cavecchia}}\email{valter.cavecchia@cnr.it}

\author*[4,5]{\fnm{Alexander} \sur{Rumyantsev}}\email{ar0@krc.karelia.ru}

\author[1,5]{\fnm{Mariia} \sur{Maltseva}}\email{mariia.maltseva@unitn.it}

\affil[1]{\orgdiv{Department of Information Engineering and Computer Science}, \orgname{University of Trento}, \orgaddress{\street{via Sommarive 9}, \city{Povo}, \postcode{38123}, \state{Trento}, \country{Italy}}}

\affil[2]{ \orgname{Trento Institute of Fundamental Physics and Applications}, \orgaddress{\street{via Sommarive 14}, \city{Povo}, \postcode{38123}, \state{Trento}, \country{Italy}}}

\affil[3]{ \orgname{Institute of Materials for Electronics and Magnetism (CNR)}, \orgaddress{\street{via alla Cascata 56/c}, \city{Povo}, \postcode{38123}, \state{Trento}, \country{Italy}}}

\affil[4]{ \orgname{Institute of Applied Mathematical Research, Karelian Research Centre of RAS}, \orgaddress{\street{11 Pushkinskaya Str.}, \city{Petrozavodsk}, \postcode{185910}, \state{Karelia Republic}, \country{Russia}}}

\affil[5]{ \orgname{Petrozavodsk State University}, \orgaddress{\street{33 Lenina Pr.}, \city{Petrozavodsk}, \postcode{185000}, \state{Karelia Republic}, \country{Russia}}}

\abstract{In this paper we introduce the Tabu Enhanced Hybrid Quantum Optimization metaheuristic approach useful for optimization 
problem solving on a quantum hardware.  We address the theoretical convergence of the proposed scheme from the viewpoint of the collisions in the object which stores the tabu states, based on the Ising model. The results of numerical evaluation of the algorithm on quantum hardware as well as on a classical semiconductor hardware model are also demonstrated.}




\maketitle

\section{Introduction}

Quantum computing (QC) is a technology of computing systems design which promises to use quantum physical phenomena to overcome the traditional limitations of sequential/parallel computing in semiconductor systems. In particular, it allows to utilize quantum state superpositions and quantum entanglement to devise algorithms that are dramatically more efficient than the classical counterparts~\cite{nielsen_quantum_2010}. A trending research subject is the study of advanced algorithms and computational problems that can successfully utilize the QC approach.

In general, the two main classes of quantum architecture are actively developed: the so-called universal QC (which can be used for general QC algorithms) and non-universal QC. One of the models of universal QC is the Adiabatic Quantum Computing (AQC). Within the AQC approach, the given problem is encoded by the problem \emph{Hamiltonian}, and the adiabatic theorem is used  to (physically) reach the minimum energy \emph{ground state} of a corresponding QC system~\cite{mcgeoch_adiabatic_2014}. In AQC the considered quantum system is assumed to be isolated so its dynamics can be described by unitary operators and hence the computations are reversible. 

The celebrated representative of the class of non-universal QC is the Quantum Annealing (QA), a physical process that can be used as an heuristic search, to solve optimization problems~\cite{wang_quantum_2016}. The solution to a given optimization problem corresponds to the \emph{ground state} of a quantum system given by a qubit lattice and described by the Ising model. The main idea is to perform a time evolution of the system towards the {ground state} in order to read out the solution of the problem by means of a measurement process. Being a rather recent concept, QA is currently under active study, both from the perspective of its hardware implementations on the available quantum computing devices~\cite{venegas-andraca_cross-disciplinary_2018} and from its theoretical and statistical properties~\cite{wang_quantum_2016}. Since in QA the quantum hardware is considered as an open system interacting with the environment, its dynamics is characterized by decoherence and energy dissipation. 

To widen the class of problems that can be solved with the help of QC, various quantum-classical hybrid schemes are introduced which use QC as one of the stages in the algorithm, e.g. by means of the so-called variational hybrid quantum-classical optimization~\cite{gentini_noise-resilient_2020} or, more widely, variational quantum algorithms (on a generic quantum hardware)~\cite{endo_hybrid_2021}, stochastic gradient descent methods~\cite{sweke_stochastic_2020} or hybrid tabu search~\cite{lee_application_2016}, to name a few. In the latter case, quantum effects are used to overcome the traps in local optima, while tabu memory improves forecasting accuracy~\cite{lee_application_2016}. In the same direction of memory-based improvement of the QC algorithm, a recent paper~\cite{pastorello2019quantum} proposed a hybrid quantum-classical algorithm based on QA to solve arbitrary Quadratic Unconstrained Optimization (QUBO) problems, called Quantum Annealing Learning Search (QALS). A further development of the idea of learning search was presented in~\cite{pastorello_learning_2021} in the context of a generic optimization problem solved by means of AQC. In the present paper, we combine these two approaches and introduce a general Tabu Enhanced Hybrid Quantum Optimization (TEHQO) metaheuristic memory-based hybrid QC approach that uses the structure of the Ising model to implement a memory mechanism.

Convergence of the algorithm is a necessary step in the analysis of any heuristics. For a non-deterministic algorithm, the usual way is to construct the corresponding stochastic process (possibly a Markov process) and observe its convergence to a steady state. In such a way, convergence of QA was established by considering inhomogeneous Markov chain in~\cite{morita_convergence_2006}, whereas various stochastic processes related notions such as hitting times and weak convergence were used in the case of the so-called quantum walks~\cite{apers_unified_2021,santos_quantum_2009,magniez_hitting_2012,balu_probability_2018}. However, in general memory comprises the theoretical convergence due to a (theoretically unlimited) dependence on the previous states of the algorithm's trajectory. Thus, it is important to  reason the finiteness of the memory of corresponding tabu search mechanism~\cite{faigle_convergence_1992,glover_tabu_2002,henderson_theory_2003,fox_integrating_1993}. We address this issue in the proposed TEHQO scheme by studying the collisions in the object storing the tabu states and using these collisions in a regenerative stochastic process representing the algorithm's execution. Therefore we suggest a novel approach to convergence of memory-based search algorithms that combines regeneration theory and algebra.

The structure of the paper is as follows. In Section~\ref{sec1} we give the necessary details on the TEHQO scheme and corresponding algorithm. In Section~\ref{sec3} we describe an approach to the formal study of convergence of the proposed algorithm based on the collisions in the tabu matrix. We give a numerical illustration of the convergence obtained from quantum hardware as well as simulation, and propose a parallel implementation of the simulation algorithm in Section~\ref{sec4}. We finalize the paper with a conclusion and discussion of further research directions.

\section{Tabu Enhanced Hybrid Quantum Optimization}\label{sec1}

In this section we introduce the TEHQO approach to solving optimization problems using QA or AQC. Before doing so, we briefly recall some generalities about AQC and QA procedures. For necessary details on AQC, QA, as well as QC in general, the reader is referred to e.g.~\cite{nielsen_quantum_2010,hughes_quantum_2021}.

AQC is a universal QC approach~\cite{AQCequiv} proposed as an application of \emph{adiabatic theorem} to solve optimization problems~\cite{Farhi2000}. The solution of a given problem corresponds to the so-called \emph{ground state} of a quantum system with total energy described by a problem \emph{Hamiltonian} $H_P$ on the Hilbert space where the considered quantum system is described. Within the mathematical formalism of quantum mechanics, the ground state corresponds to the eigenvector of $H_P$ with minimum eigenvalue (if the eigenspace with lowest eigenvalue is multidimensional, the ground state is called \emph{degenerate}). 

The informal structure of an AQC is the following: the given problem is encoded in the problem Hamiltonian $H_P$ and the quantum system is prepared in the known ground state of an initial Hamiltonian $H_I$. Then a time variation of the Hamiltonian from $H_I$ to $H_P$ is implemented according to
\begin{equation}
H(t)=(1-s(t))H_I+s(t)H_P \quad t\in[0,\tau],
\end{equation}
where $s:[0,\tau]\rightarrow[0,1]$ is a smooth monotone function such that $s(0)=0$ and $s(\tau)=1$. According to the adiabatic theorem, if the change is sufficiently slow then the system remains in the instantaneous ground state with high probability. The time evolution must be slow enough to satisfy the adiabatic condition without destroying the efficiency of the QC~\cite{adiabatic_condition}. The solution is then obtained as the ground state of the system at time $\tau$.

QA is a physical process that can be used as an heuristic search to solve optimization problems~\citep{PhysRevE.58.5355}.   
The QA procedure is implemented by a time evolution (the \emph{cooling}) of the quantum system, characterized by the energy dissipation and decoherence, towards the ground state of the problem Hamiltonian $H_P$. QA is closely related to AQC since the solution of the problem is encoded into the ground state of a problem Hamiltonian, however the considered quantum system is not isolated and the evolution is non-unitary so QA does not simulate universal QC in general. 

There is a conceptual analogy between QA and classical Simulated Annealing (SA). The latter is a local search optimization heuristic, a representative of the so-called Generalized Hill Climbing class of algorithms~\cite{johnson_convergence_2002}, that resembles the annealing physical process in the metal production. The remarkable difference between QA and SA is that QA exploits the \emph{tunnel effect}, a purely quantum phenomenon, to escape local minima instead of thermal hill-climbing, which, in turn, makes QA indeed a \emph{global} search. 

\begin{algorithm}
\caption{Tabu Enhanced Hybrid Quantum Optimization}\label{Schema}
\begin{algorithmic}[1]
\Require Function $f:\mathcal{Z}\rightarrow \mathbb R$ to be minimized on $\mathcal{Z}=\{-1,1\}^n$
\Require maximum temperature $T_{max}$, function $h$ for temperature decreasing depending on temperature decreasing rate $\eta>0$
\Require parametric Hamiltonian $H^{(\boldsymbol\alpha)}$, randomized initialization function $\zeta$ and randomized temperature-dependent modification function $\xi_T$ for the parameters $\boldsymbol\alpha$, both functions depending explicitly on $f$
\Require number $N$ of iterations with constant temperature
\Require probability $q$ of performing ground state measurement
\Require maximum number of iterations $N_{max}$ and $i_{max}$
\Require QC procedure to obtain a ground state $\boldsymbol z$ of the tabu-enhanced parametric Hamiltonian $H_P^{(\boldsymbol\alpha)}(\boldsymbol S)$, say, QA or AQC
\Ensure $\boldsymbol z^{\star}\in\mathcal{Z}$ vector minimum of $f$
\State $T\gets T_{max}, \boldsymbol S\gets\boldsymbol{O}$
\State initialize $\boldsymbol\alpha_1 \gets \zeta(f)$ and  $\boldsymbol\alpha_2\gets \zeta(f)$
\State initialize Hamiltonians $H_P^{(\boldsymbol\alpha_1)}(\boldsymbol O)$ and $H_P^{(\boldsymbol\alpha)}(\boldsymbol O)$ as in~\eqref{H(alpha,theta)}
\State find $\boldsymbol z_1$ and $\boldsymbol z_2$ by QC procedure
\State evaluate $f(\boldsymbol z_1)$ and $f(\boldsymbol z_2)$
\If{$f(\boldsymbol z_1)\neq f(\boldsymbol z_2)$}
  \State use the best to initialize $\boldsymbol z^{\star}$\ and the corresponding parameters $\boldsymbol\alpha^\star$.
  \State use the worst to initialize $\boldsymbol z'$
  \State initialize the tabu matrix: $\boldsymbol S \gets \boldsymbol m( \boldsymbol z')$
\EndIf
\State $e\gets 0, d\gets 0, i\gets 0$
\Repeat
\If{$N$ divides $i$}\Comment{temperature decrease}
    \State{$T\gets h(\eta,T)$}
\EndIf
\State with probability $q$ set $\boldsymbol\alpha\gets \xi_T(\boldsymbol\alpha^\star,f)$, initialize the Hamiltonian $H_P^{(\boldsymbol\alpha)}(\boldsymbol S)$ using~\eqref{H(alpha,theta)} and find $\boldsymbol z'$ by QC procedure; otherwise sample $\boldsymbol z'$ using quantum random number generator
\If{$\boldsymbol z' \neq \boldsymbol z^{\star}$}
\State evaluate $f(\boldsymbol z')$
\If{$f(\boldsymbol z')<f(\boldsymbol z^{\star})$}\Comment{new candidate solution}
\State $swap(\boldsymbol z',\boldsymbol z^{\star})$,  $\boldsymbol\alpha^\star \gets \boldsymbol\alpha$, $e\gets 0, d\gets 0$
\Else\Comment{suboptimal acceptance}
\State $d\gets d+1$
\State with probability $e^{-\frac{f(\boldsymbol z')-f(\boldsymbol z^{\star})}{T}}$ $swap(\boldsymbol z',\boldsymbol z^{\star})$, $\boldsymbol\alpha^\star \gets \boldsymbol\alpha$, $e\gets 0$
\EndIf
\State $\boldsymbol S\gets \boldsymbol S+\boldsymbol m(\boldsymbol z')$ \Comment{tabu matrix update}
\Else
\State $e\gets e+1$
\EndIf
\State $i\gets i+1$
\Until{$i=i_{max}$ or $d+e>N_{max}$}
\State \textbf{return} $\boldsymbol z^{\star}$
\end{algorithmic}
\end{algorithm}

QA can be physically realized considering a quantum spin glass that is a network of qubits arranged on the vertices of a graph $\langle V,E\rangle$, with $\lvert{V}\rvert=n$, whose edges in $E$ represent the couplings among the qubits. The total energy of such a system is represented by the Hamiltonian $H_P\equiv H_{Ising}(\boldsymbol\Theta)$ of the quantum spin glass,
\begin{equation}\label{HP}
H_{Ising}(\boldsymbol\Theta)=\sum_{i\in V} \theta _i  \sigma^{(i)}_3 +\sum_{(i,j)\in E} \theta_{ij} \sigma^{(i)}_3 \sigma^{(j)}_3.
\end{equation}
$H_{Ising}(\boldsymbol\Theta)$ is an operator on the $n$-qubit Hilbert space $\mathsf H=(\mathbb C^2)^{\otimes n}$ where the $i$th qubit term is represented by a tensor product, 
\[
\sigma^{(i)}_3:=\boldsymbol I_{2^{i-1}}\otimes  \sigma_3\otimes\boldsymbol I_{2^{n-i}},\quad i=1,\dots,n,
\]
where $\boldsymbol I_k$ is the identity matrix of size $k$ and $\sigma_3$ is the so-called Pauli-Z matrix placed in the $i$th tensor factor:
\[
\sigma_3=\begin{bmatrix}
1 & 0\\
0 & -1
\end{bmatrix},
\]
whereas the interaction of $i$th and $j$th qubits is represented by the term
\[
 \sigma^{(i)}_3 \sigma^{(j)}_3=\boldsymbol I_{2^{i-1}}\otimes  \sigma_3\otimes\boldsymbol I_{2^{j-i-1}}\otimes  \sigma_3\otimes\boldsymbol I_{2^{n-j}},\quad j>i=1,\dots,n.
\]
The coefficient matrix $\boldsymbol\Theta$ is the symmetric square size-$n$ matrix of the real coefficients of the Hamiltonian $H_{Ising}$, called \emph{weights}, defined as
\begin{equation}\label{w}
\boldsymbol\Theta=||\theta_{ij}||_{(i,j)\in V\times V}:=\left\{
\begin{array}{ll}
\theta_i, & i=j,\\
\theta_{ij}, & (i,j)\in E,\\
0, & (i,j)\not\in E.
\end{array}\right.
\end{equation}
These coefficients physically correspond to the coupling terms between the qubits, $\theta_{ij}$, and the so-called local fields, $\theta_i$, on the corresponding vertices.

The matrix $\sigma_3$ has two eigenvalues corresponding to the binary states of the qubit, $\{-1,1\}$, and thus, the system~\eqref{HP} has the spectrum of eigenvalues corresponding to all possible values of the cost function given by the energy of the well-known \emph{Ising model}:
\begin{equation*}
    \mathsf E(\boldsymbol{\Theta}, \boldsymbol z)=\sum_{i\in V} \theta_i z_i +\sum_{(i,j)\in E} \theta_{ij}z_i z_j,\; \boldsymbol z=(z_i)_{i\in V}\in\mathcal{Z}=\{-1,1\}^n.
\end{equation*}
Thus, the annealing procedure takes the system to the ground state whose corresponding spin configuration is
\begin{equation}\label{argmin_zE}
\boldsymbol z^{\star}=\arg\min_{\boldsymbol z\in\mathcal{Z}} \mathsf E(\boldsymbol\Theta,\boldsymbol z),
\end{equation}
where $\boldsymbol z^{\star}$ is the optimal solution of the optimization problem encoded into the parameters of the Hamiltonian $H_{Ising}(\boldsymbol \Theta)$ defined in~\eqref{HP}.

Given a problem, the QA is initialized by a suitable choice of the weights $\boldsymbol\Theta$ and the binary variables $z_i\in\{-1,1\}$ are physically realized by the outcomes of measurements on the qubits located in the vertices $V$. Thus, since $\boldsymbol\Theta$ is symmetric, the optimization problem is specified by $n(n-1)/2$ real parameters satisfying some architectural constraints of a quantum machine, 
\[
\theta_{ij}\in[-\Delta, +\Delta],
\] 
for all $(i,j)\in E$ and some finite positive $\Delta$, say, 1. In order to solve a general optimization problem by QA, one needs to obtain the correct \textit{encoding} of the objective function in terms of the cost function $\mathsf E(\boldsymbol{\Theta}, \boldsymbol z)$ which is hard in general. 

Now we present the TEHQO scheme, a guided meta-heuristic approach designed specifically to solve optimization problems without the need of representing a problem into the quantum architecture a priori. This idea generalizes both the QALS approach to solve QUBO problems using QA, suggested in~\citep{pastorello2019quantum}, and AQCLS approach applicable in AQC environment, described in~\citep{pastorello_learning_2021}. Below we address this generalization by explaining the key idea in QALS and inspiration from AQCLS.

The key idea of the QALS is to enhance the QA by adding the so-called \emph{tabu matrix} $\boldsymbol S$ to the weight matrix $\boldsymbol\Theta$ of the coefficients of $H_{Ising}$ so as to penalize the solutions already visited and prevent a redundant search in the solution space. The matrix is constructed in additive way using the set of $k$ already visited (worst) solutions $\{\boldsymbol z_j\}_{j=1,...,k}$ with the help of a function 
\begin{equation}\label{constructtabu}
   \boldsymbol m(\boldsymbol z)=\boldsymbol z \boldsymbol z^T-\boldsymbol I_n+\mathrm{diag} (\boldsymbol z),\quad \boldsymbol z\in\mathcal{Z},
\end{equation}
where $\mathrm{diag}(\boldsymbol z)$ constructs a diagonal matrix from a vector $\boldsymbol z$. Note that, by construction, the matrix $\boldsymbol m(\boldsymbol z)$ is symmetric and, moreover,
\begin{equation}\label{penalty}
\mathsf E(\boldsymbol m(\boldsymbol z),\boldsymbol z)>0,\quad \boldsymbol z\in\mathcal{Z}.
\end{equation}
The matrix $\boldsymbol S$ is then constructed as the sum:
\begin{equation}\label{matrix}
    \boldsymbol S=\sum_{j=1}^k \boldsymbol m(\boldsymbol z_j).
\end{equation}
Due to~\eqref{penalty}, the tabu matrix $\boldsymbol{S}$ introduces \emph{energetic penalties} on the solutions $\{\boldsymbol z_j\}_{j=1,...,k}$ in the spectrum of the Hamiltonian $H_{Ising}(\boldsymbol\Theta+\boldsymbol S)$. 

Following the key idea of QALS, the TEHQO scheme is based on an iterative procedure of candidate solutions generation by reaching and measuring the ground state (by QA or AQC) of a parameter dependent problem Hamiltonian of the form 
\begin{equation}\label{H(alpha,theta)}
H_P^{(\boldsymbol\alpha)}(\boldsymbol S)=H^{(\boldsymbol\alpha)}+H_{Ising}(\boldsymbol S),
\end{equation}
where $H^{(\boldsymbol\alpha)}$ is a generic (arbitrary) Hamiltonian depending on the parameters $\boldsymbol\alpha$ which are updated along with the candidate solutions and $H_{Ising}$ is the Ising Hamiltonian used to enable the solution penalties by means of the tabu matrix $\boldsymbol S$. From the physical point of view, we can have a spin glass described by an Ising Hamiltonian on which we can act with controlled external fields in order to implement $H^{(\boldsymbol\alpha)}$. As a particular case, if
\begin{equation}
H^{(\boldsymbol \alpha)}=H_{Ising}(\boldsymbol\alpha),
\end{equation}
then we recover the QALS scheme and \[H_P^{(\boldsymbol\alpha)}(\boldsymbol S)=H_{Ising}(\boldsymbol\alpha+\boldsymbol S).
\] 
In such a case, $\boldsymbol \alpha\equiv \boldsymbol \Theta$ where $\boldsymbol\Theta$ is as given in~\eqref{w}. 

Another example inspired by~\cite{pastorello_learning_2021} is  $H^{(\boldsymbol\alpha)}$ realized by means of the transverse fields that presents the following form:
\begin{equation}
H^{(\boldsymbol\alpha)} =\sum_{i\in V_1} \alpha_i  \sigma^{(i)}_1 +\sum_{(i,j)\in E_1} \alpha_{ij} \sigma^{(i)}_1 \sigma^{(j)}_1,
\end{equation}
where $\sigma_1$ is the Pauli-X matrix
\[
\sigma_1=\begin{bmatrix}
0 & 1\\
1 & 0
\end{bmatrix},
\]
and the graph $(V_1,E_1)$ can differ from the graph $(V,E)$ of $H_{Ising}$. Generally speaking, we do not require any specific constrain over the form of $H^{(\boldsymbol\alpha)}$ that represents the class of Hamiltonians which can be physically implemented in a laboratory or can be efficiently simulated by suitable quantum circuits. As such, compared to QALS, the TEHQO scheme is not limited to QUBO problems, but it designed to solve arbitrary optimization problems considering a more general quantum architecture. 

Within TEHQO scheme, the search is simultaneously performed among the candidate solutions $\boldsymbol z$ (to find the optimum) and in the space of the parameters $\boldsymbol\alpha$ of the Hamiltonian $H^{(\boldsymbol\alpha)}$ (to find the best representation of the problem in the space of  parametrized Hamiltonians). The candidate solutions are obtained by means of the QC part, and suboptimal acceptance of the candidate solutions is allowed at the classical counterpart of the algorithm, so the classical part of the hybrid algorithm presents a SA-like structure. The rejected solutions are used to update the tabu matrix $\boldsymbol{S}$ and new candidate solutions are generated, while the parameters $\boldsymbol\alpha$ of the Hamiltonian $H^{(\boldsymbol\alpha)}$ are perturbed in temperature-dependent way, with decreasing temperature parameter. In this work we consider a randomized temperature-dependent function $\xi_T$ to modify the Hamiltonian parameters within the iterative structure of the proposed algorithm, which again is a generalization of the QALS scheme.

To finalize the comparison, we note that TEHQO provides a generalization of a specific AQCLS. In AQCLS the search within the family of available Hamiltonians is carried on randomly by Gaussian sampling of the parameters. Instead TEHQO considers a deformation function over the parameters which takes into the account the considered problem by the explicit dependence of $\xi_T$ on $f$. However, TEHQO presents a tabu Hamiltonian of Ising-type defined by means of the tabu matrix, instead AQCLS provides a general Hamiltonian defined by a tabu list.

The TEHQO scheme is presented in Algorithm~\ref{Schema}.
The QC procedure (say, QA or AQC) is run with parameters that are initialized by the randomized function $\zeta$ depending on the objective function $f$ of the problem (lines 2, 3, 4), and the worst solution $\boldsymbol z'$ is used to initialize the tabu matrix according to~\eqref{matrix} (line 9). In the iterative structure, the temperature $T$ is constant for a cycle of $N$ iterations (line 13) and decreases according to the function $h$ and the rate $\eta$ at any $N$-th cycle (line 14). For instance, if $h(\eta,T)=T-\eta$ then $T$ decreases linearly, if $h(\eta,T)=T-\eta T$ then $T$ decreases exponentially. At each iteration (line 16) one of the two alternatives is taken, either (with probability $q$) new parameters $\boldsymbol\alpha$ are generated perturbing $\boldsymbol\alpha^\star$ corresponding to the current best candidate solution using a temperature-dependent modification function $\xi_T$ which explicitly depends on $f$, or (with probability $1-q$) $\boldsymbol z'$ is sampled using a quantum random number generator, e.g. by measurement on the quantum superposed state. In the former case, the Hamiltonian $H_P^{(\boldsymbol\alpha)}(\boldsymbol S)$ is initialized and the QC procedure produces the new candidate solution $\boldsymbol z'$ (line 16) that is tested (line 18-19). If $\boldsymbol z'$ is better then it is accepted and the parameters are updated accordingly (line 20), else $\boldsymbol z'$ is rejected or accepted as a suboptimal solution using SA-like scheme (line 23). Consequently, the tabu matrix is updated in either case (line 25). The Algorithm~\ref{Schema} terminates either if the maximum number $i_{max}$ of iterations is achieved, or the convergence to a solution of the optimization problem is stated (line 30).

Since the TEHQO is iteratively applied, the convergence of the procedure to the optimal solution $\boldsymbol z^{\star}$ is usually proved by considering the convergence of the sequence of steady-state probabilities of non-homogeneous Markov chains modeling the sequences of candidate solutions obtained. However, due to~\eqref{matrix}, proving the Markovian nature of such a process is problematic, since the tabu matrix holds virtually unlimited memory of previous states. Thus, it is important to study the collisions in the tabu matrix $\boldsymbol S$. We address this issue in the next section.

\section{Theoretical convergence}\label{sec3}

The convergence proof of the TEHQO scheme based on some properties of the SA-like structure implemented by the classical part of Algorithm~\ref{Schema}. In order to prove the algorithm convergence we consider the stochastic sequence of the candidate solutions obtained in the algorithm run. At the same time, we need to show finiteness of memory of the tabu matrix $\boldsymbol S$.

In order to define the tabu matrices corresponding to the run of TEHQO algorithm, let us enumerate the state space. Let the $n$-dimensional column vector $\boldsymbol z^{(i)}\in \mathcal{Z}$ be $i$th element of the lexicographically ordered state space $\mathcal{Z}$ (this may be constructively defined as the $n$-digit binary representation of $i$ in the binary alphabet $\{-1,1\}$).

Consider now two independent runs of TEHQO algorithm. For each such a run encode the sequence of rejected solutions by a nonnegative integer sequence of length $2^n$ (irregardless of the order in which they appeared in the trajectory). For each such a sequence $\boldsymbol a\in \mathbb{Z}_+^{2^n}$ construct the matrix function $\boldsymbol S(\boldsymbol a)$ using~\eqref{matrix} and~\eqref{constructtabu} in the following way:
\begin{equation}\label{smatrix}
    \boldsymbol S(\boldsymbol a)=\sum_{i=1}^{2^n} a_i\; \boldsymbol m\left(\boldsymbol z^{(i)}\right).
\end{equation}
The \textit{collision} in the tabu matrix happens if $\boldsymbol S(\boldsymbol a)=\boldsymbol S(\boldsymbol b)$ for $\boldsymbol a\neq \boldsymbol b\in \mathbb{Z}_+^{2^n}$. Such collisions are solutions of a linear system
\begin{equation}\label{whattodo2}
    \sum_{i=1}^{2^n} x_i\; \boldsymbol m\left(\boldsymbol z^{(i)}\right)=\boldsymbol{O},
\end{equation}
where $\boldsymbol{O}$ is the square zero matrix of dimension $n$ and $x_i:=a_i-b_i$ are the variables. Rewriting~\eqref{whattodo2} in a vector-matrix form, obtain
\begin{equation}\label{whattodo}
    \boldsymbol M \boldsymbol x=\boldsymbol{0},
\end{equation}
where $\boldsymbol x\in \mathbb{Z}^{2^n}$ is the (column) vector of componentwise differences of two trajectories and $\boldsymbol M$ is the $\frac{n(n+1)}{2} \times 2^n$ matrix containing columnwise the upper-triangular elements of matrices $\boldsymbol m(\boldsymbol z^{(i)})$, starting from the diagonal elements.  As such, solution of the system~\eqref{whattodo} is a vector in the kernel of~\eqref{smatrix}. In particular, if $\boldsymbol x\in \mathbb{Z}_+^{2^n}$, then a sequence of states corresponding to $\boldsymbol x$ produces a zero tabu matrix after a non-zero number of iterations of the QA algorithmn, i.e. $\boldsymbol S(\boldsymbol x)=\boldsymbol{O}$. Let us formally study the solutions of the system~\eqref{whattodo}.

Now we recursively define a sequence of matrices, which allows us to establish the desired algebraic properties. Define the following $2^k\times 2^{k-1}$ matrices for any $k\geq 1$:
\begin{equation}\label{defSA}
\mathbb S^{(k)}:=\begin{bmatrix}
\mathbb I_{2^{k-1}}\\
\mathbb J_{2^{k-1}}
\end{bmatrix},
\qquad
\mathbb A^{(k)}:=\begin{bmatrix}
-\mathbb I_{2^{k-1}}\\
\mathbb J_{2^{k-1}}
\end{bmatrix},
\end{equation}
where $\mathbb I_{2^{k-1}}$ and $\mathbb J_{2^{k-1}}$ are the identity matrix and the exchange matrix (matrix with unit vector over antidiagonal) of order $2^{k-1}$ respectively (conventionally $\mathbb I_0=\mathbb J_0=1$). Construct the class $SA(n)$ of $2^n\times 1$ matrices having the following form 
\[
F^{(n)}F^{(n-1)}\cdots F^{(1)}\in SA(n),
\]
where $F^{(k)}$ can be either $\mathbb S^{(k)}$ or $\mathbb A^{(k)}$. In the following proposition we establish group properties of $SA(n)$.
\begin{proposition}{}\label{ortho}
The following properties of $SA(n)$ are true for any $n\geqslant 1$:
\begin{enumerate}
\item With the exception of $\mathbb S^{(n)}\cdots \mathbb S^{(1)}$, any element of $SA(n)$ contains the same number of $-1$ and $1$, equal to $2^{n-1}$.
\item $SA(n)$ equipped with the Hadamard (componentwise) product $\circ$ is an abelian group with the neutral element $\mathbb S^{(n)}\cdots \mathbb S^{(1)}$ and any element being the inverse of itself.
\item The elements of $SA(n)$ are orthogonal w.r.t. the Euclidean scalar product.
\end{enumerate} 
\end{proposition}

\begin{proof}{} In the proof, we use induction on $n$.

(1) For $n=1$ the group $SA(1)$ contains only $\mathbb S^{(1)}$ and $\mathbb A^{(1)}=\begin{bmatrix}-1\\1\end{bmatrix}$. Let (1) hold good for $SA(j)$, $j\leq n$. Denote 
\[
v=F^{(n)}F^{(n-1)}\cdots F^{(1)},
\]
where $F^{(k)}\in\{\mathbb S^{(k)}, \mathbb A^{(k)}\}$ for $k\leq n+1$. Then since
\[
\mathbb S^{(n+1)}v=\begin{bmatrix}
v\\
\mathbb J_{2^n} v
\end{bmatrix} \mbox{ and }
\mathbb A^{(n+1)}v=\begin{bmatrix}
-v\\
\mathbb J_{2^n} v
\end{bmatrix},
\]
the statement (1) follows for $n+1$ and the induction step follows.

(2) For $n=1$, we have 
\[
\mathbb S^{(1)}\circ \mathbb S^{(1)}, \mathbb S^{(1)}\circ \mathbb A^{(1)}, \mathbb A^{(1)}\circ \mathbb S^{(1)}, \mathbb A^{(1)}\circ \mathbb A^{(1)} \in SA(1),
\] 
by definition~\eqref{defSA}. Let (2) hold good for $SA(j)$, $j\leq n$. Denote 
\[
v=F^{(n)}F^{(n-1)}\cdots F^{(1)},\;  w=E^{(n)}E^{(n-1)}\cdots E^{(1)},
\]
where $E^{(k)}, F^{(k)}\in\{\mathbb S^{(k)}, \mathbb A^{(k)}\}$, $k\leq n+1$. Similarly to the proof of statement (1),
\begin{equation}
F^{(n+1)}v=\begin{bmatrix}
\pm v\\
\mathbb J_{2^n} v
\end{bmatrix},
\qquad
E^{(n+1)}w=\begin{bmatrix}
\pm w\\
\mathbb J_{2^n} w
\end{bmatrix},
\end{equation}
and hence
$$   F^{(n+1)}v\circ E^{(n+1)}w=\begin{bmatrix}
\pm v\circ w\\
\mathbb J_{2^n} (v\circ w)
\end{bmatrix} \in SA(n+1)$$ holds by definition of $SA(n+1)$, since 
\[\begin{bmatrix}
 v\circ w\\
\mathbb J_{2^n} (v\circ w)
\end{bmatrix}= \mathbb S^{(n+1)} (v\circ w),
\]
\[\begin{bmatrix}
- v\circ w\\
\mathbb J_{2^n} (v\circ w)
\end{bmatrix}= \mathbb A^{(n+1)} (v\circ w),
\]
and finally $v\circ w\in SA(n)$, which completes the induction step. It remains to note that the neutral element is given by $\mathbb S^{(n)}\cdots \mathbb S^{(1)}$, whose entries are identically $1$, as a direct consequence of the definition of Hadamard product, and hence each element of the group is inverse of itself.

(3) As a  consequence of (1) and (2) we have that the Euclidean scalar product between two distinct elements $v$ and $w$ of $SA(n)$ is nothing but the sum of the entries of $v\circ w$ that presents an equal number of $-1$s and $1$s. 
\end{proof}
\medskip

Now we consider the matrix $\boldsymbol M$ of the linear system~\eqref{whattodo} and prove the following statement.

\begin{proposition}{}\label{Mrows}
The rows of $\boldsymbol M$ are transposed elements of a subset $SA_M\subset SA(n)$.
\end{proposition}
\begin{proof}{}
Construct a $2^n\times n$ matrix $\boldsymbol B$ by columns $B_i$, $i=1,\dots,n$, using specific vectors from the $SA(n)$ group as follows:
\[
B_i=F^{(n)}\dots F^{(1)}, \quad F^{(i)}=\mathbb A^{(i)}, F^{(j)}=\mathbb S^{(j)}, j\neq i.
\]
This means that $\boldsymbol B$ by column consists of the column vectors $\mathbb A^{(n)} \mathbb S^{(n-1)}\dots \mathbb S^{(1)}$, $\mathbb S^{(n)} \mathbb A^{(n-1)}\dots \mathbb S^{(1)}$ and so on. However, this process can be done recursively starting from the last, $n$th column, as follows. We start with the column vector $\mathbb A^{(1)}=\begin{bmatrix}-1\\1\end{bmatrix}$ of two components. Then we left-multiply it with $\mathbb S^{(2)}$ which in fact produces a four-component vector containing the values of $\mathbb A^{(1)}$ mirrored, i.e. $(-1, 1, 1, -1)^T$. Now we add from the left another column (which finally becomes the column $n-1$ in the constructed matrix), which contains values $\mathbb A^{(2)} \mathbb S^{(1)}=(-1, -1, 1, 1)^T$. This means that the original column $\mathbb A^{(1)}$ corresponds to the values $-1$ of the column added from the left, whereas the mirrored column $\mathbb A^{(1)}$ corresponds to the values $1$ in the left column. These two actions (mirroring, adding from the left a column vector of sequence of $-1$'s followed by $1$'s of equal length) is repeated until the constructed matrix $\boldsymbol B$ has $n$ columns. 

It is easy to see that the resulting matrix $\boldsymbol B$ contains classical Gray (binary reflected) codes in the binary alphabet $\{-1,1\}$, according to recursive Gray code generation procedure, see e.g.~\citep{knuthart}. It is well known that the Gray codes of length $n$ enumerate all the possible values of $2^n$ strings in the binary alphabet. As such, there exists a permutation 
\[
p: \{1,\dots,2^n\}\mapsto \{1,\dots,2^n\},
\]
such that $p(i)$ is the row number in the matrix $\boldsymbol B$ of the (classical) binary representation of the value $i$ in the alphabet $\{-1,1\}$ having length $n$. From this matrix, we construct an $\frac{n(n+1)}{2} \times 2^n$ matrix $\boldsymbol A$ from the transposed matrix $\boldsymbol B^T$ followed by pairwise Hadamard products of its rows, 
\[
B_i\circ B_j, i<j, i,j\in\{1,\dots,n\}.
\]
Note that, by statement (2) of Proposition~\ref{ortho}, all rows of $\boldsymbol A$ are transposed elements of $SA(n)$. We denote these elements as a subset 
\[
SA_M\subset SA(n).
\]

Recall now the construction of matrix $\boldsymbol M$. It contains by columns all the values of diagonal and upper triangular elements of matrices $\boldsymbol m(\boldsymbol z^{(i)})$ using~\eqref{constructtabu}, where $\boldsymbol z^{(i)}$ is the length-$n$ binary representation of $i=1,\dots,2^n$ in the alphabet $\{-1,1\}$. As such, $i$th column of the matrix $\boldsymbol M$ corresponds to $p(i)$th column of matrix $\boldsymbol A$, which completes the proof.
\end{proof}
\medskip

An immediate consequence of Propositions~\ref{ortho} and~\ref{Mrows} is the following Proposition.
\begin{proposition}{}\label{main}
The space of solutions $\mathcal V_M$  of the linear system~\eqref{whattodo} is spanned by the elements of $SA(n)\setminus SA_M$ and its dimension is:
$$\dim\mathcal V_M=2^n-\frac{n(n+1)}{2}.$$
\end{proposition}
Intuitively this means that any solution $\boldsymbol x$ of the system~\eqref{whattodo} is a linear combination of the vectors in $SA(n)\setminus SA_M$ which form a basis being orthogonal w.r.t. Euclidean scalar product, according to (3) of Proposition~\ref{ortho}. At the same time, if the solution vector is non-negative, $\boldsymbol x\geq \boldsymbol 0$, then the corresponding tabu matrix $\boldsymbol S(\boldsymbol x)$, constructed according to~\eqref{smatrix}, will be a zero matrix. Below we use this result to prove the convergence of TEHQO algorithm.

In Section~\ref{sec1}, we described the candidate solution generation by Algorithm~\ref{Schema}. In what follows, we study the convergence of the TEHQO trajectory under two simplifying assumptions:
\begin{enumerate}
    \item the sequence of candidate solutions forms a non-homogeneous temperature-dependent stochastic sequence in a finite state space $\mathcal{Z}$;
    \item the sequence of tabu matrices evolves on a finite (multidimensional) state space $\mathcal{S}$.
\end{enumerate}
The latter assumption may be relaxed, however, it causes additional efforts to prove positive recurrence of the corresponding two-dimensional process describing the TEHQO algorithm evaluation.

We consider the two-dimensional discrete-time stochastic process \begin{equation}\label{twodimprocess}
\{\boldsymbol Z_i^{(T)}, \boldsymbol S_i^{(T)}\}_{i\geq 1}
\end{equation} living in the state space $\mathcal{Z}\times\mathcal{S}$, where $\boldsymbol Z_i^{(T)}$ is the current candidate solution and $\boldsymbol S_i^{(T)}$ the tabu matrix at $i$th step of the algorithm. We note that the dependence on the current temperature, $T$, is stressed in the notation, however, the sequence of temperatures is a non-random sequence $\{T_j\}_{j\geq 1}$ which is selected in a non-random way. Now we show that the process~\eqref{twodimprocess} is regenerative.

\begin{proposition}{}
Algorithm~\ref{Schema} converges.\label{prop1}
\end{proposition}

\begin{proof}{}
Since the component $\boldsymbol Z_i^{(T)}$ of the process~\eqref{twodimprocess} lives in a finite state space $\mathcal{Z}$, we can build a sequence of random times $t_1,t_2,\dots$ such that $\boldsymbol Z_i^{(T)}$ visits a specific state, say, $\hat{\boldsymbol z} \in\mathcal{Z}$. Due to the quantum nature of the computation, the probability of generating the candidate $\hat{\boldsymbol z}$ given any current solution $\boldsymbol z'\neq \hat{\boldsymbol z}$ is uniformly bounded from below by the value ${(1-q)}/{2^n}$, where $1-q$ is the probability to obtain the state by quantum random number generator (line 16 of Algorithm~\ref{Schema}), in this case the outcome is produced by a uniform sampling on $\mathcal{Z}$. As such, since $\mathcal{Z}$ is finite, geometrical trial argument (see e.g.~\citep{asmussen_applied_2003}) allows to conclude that the sequence $\{t_j\}_{j\geq 1}$ is infinite and has finite mean inter-event times.

Using Proposition~\ref{main} we take a subsequence $\{t_{j_k}\}_{k\geq 1}$ such that $\boldsymbol S^{(T)}_{t_{j_k}}=\boldsymbol{O}$. The probability of obtaining such a matrix from any other matrix, say, $S'$, is bounded from below by positive value. Indeed, due to the fact that there are many solutions of the system~\eqref{whattodo}, we select the closest non-negative solution $\boldsymbol x\geq \boldsymbol 0$ such that $\boldsymbol x\geq \boldsymbol y$, where $\boldsymbol y$ is any trajectory that gives a tabu matrix $S'$. Due to assumption of finiteness of the state space $\mathcal{S}$, the maximal distance between the vectors $\boldsymbol x$ and $\boldsymbol y$, that is, $\max_{k=1,\dots,2^n} \lvert x_k-y_k\rvert$, is finite. Thus, the number of steps to reach $\boldsymbol x$ from $\boldsymbol y$ by adding the solutions to tabu matrix, is uniformly bounded from above. Finally, we conclude that the matrix $\boldsymbol S^{(T)}_{t_{j_k}}$ regenerates at zero with positive probability, and using the geometrical trial argument, conclude that the subsequence $\{t_{j_k}\}_{k\geq 1}$ has finite mean. Thus, overall the process~\eqref{twodimprocess} is a positive recurrent regenerative process which guarantees the convergence of Algorithm~\ref{Schema}.
\end{proof}

\section{Numerical Illustration}\label{sec4}

In this section we numerically illustrate the theoretical results related to the convergence of TEHQO scheme. Namely, we use a simplified version of the QALS algorithm, which is a particular case of the TEHQO scheme, to study the collisions in the tabu matrix. We do so by solving a simple QUBO problem, firstly by performing simulation and secondly by running the minimization in hybrid regime on QA hardware. In both experiments, we watch the tabu matrix and observe the iteration at which the matrix first becomes zero.

The specific QUBO problem was the minimization of the quadratic function
$f(\boldsymbol x)=\boldsymbol x\boldsymbol Q\boldsymbol x$ having the following matrix
\[
\boldsymbol Q=\begin{bmatrix}
0& -0.2& 0& 0\\
0& 1.0& -0.5& 1.0\\
0& 0& -0.9& 0\\
0& 0& 0& 0.7\\
\end{bmatrix}.
\]
This results in the following function to be minimized:
\[
f(\boldsymbol x)=-0.2 x_1 x_2+ x_2^2-0.5 x_2 x_3+ x_2 x_4 -0.9 x_3^2 +0.7 x_4^2,
\]
where the variables $x_i\in \{-1,1\}, i=1,\dots,4$. This function has minima at $\boldsymbol x=(-1, -1, -1, 1)$ and, symmetrically, at $-\boldsymbol x$, with the value of $-0.9$.

We use the following configuration of the parameters related to Algorithm 1:
\[
i_{max}=200,\; N_{max}=100,\; q=0.99,\; \eta=0.2.
\]
In the considered version of QALS (introduced in~\cite{pastorello2019quantum}), the role of temperature parameter is played by a sequence of probabilities $\{p_i\}_{i\geq 1}$ that decrease geometrically with $i$ from $p_1=1$ to the fixed value $p_{\delta}=0.01$, and the temperature parameter $T$ is related to (a generic member of the sequence) $p$ by the following relation 
\begin{equation}\label{relationshippdelta}
p-p_{\delta}=e^{-\frac{1}{T}}.
\end{equation}
Since the recursive relation to obtain the sequence of probabilities $\{p_i\}_{i\geq 1}$ in~\cite{pastorello2019quantum} was 
\[
p_{i+1} = p_i-(p_i-p_{\delta})\eta,\quad, i\geq 1,
\] 
the temperature modification function for the algorithm can be easily deduced from~\eqref{relationshippdelta} as
\[
h(\eta,T)=\frac{T}{1-T\log(1-\eta)}.
\]
Moreover, due to~\eqref{relationshippdelta}, the initial value $T_{max}$ is given as
\[
T_{max}=-\left(\log(0.99)\right)^{-1}.
\]
Finally, the initialization $\zeta$ was a uniform (in $[0,1]$) random number generator and the modification $\xi_T$ was the identity function.

\begin{figure}[!b]
  \centering
  \includegraphics[width=.9\textwidth]{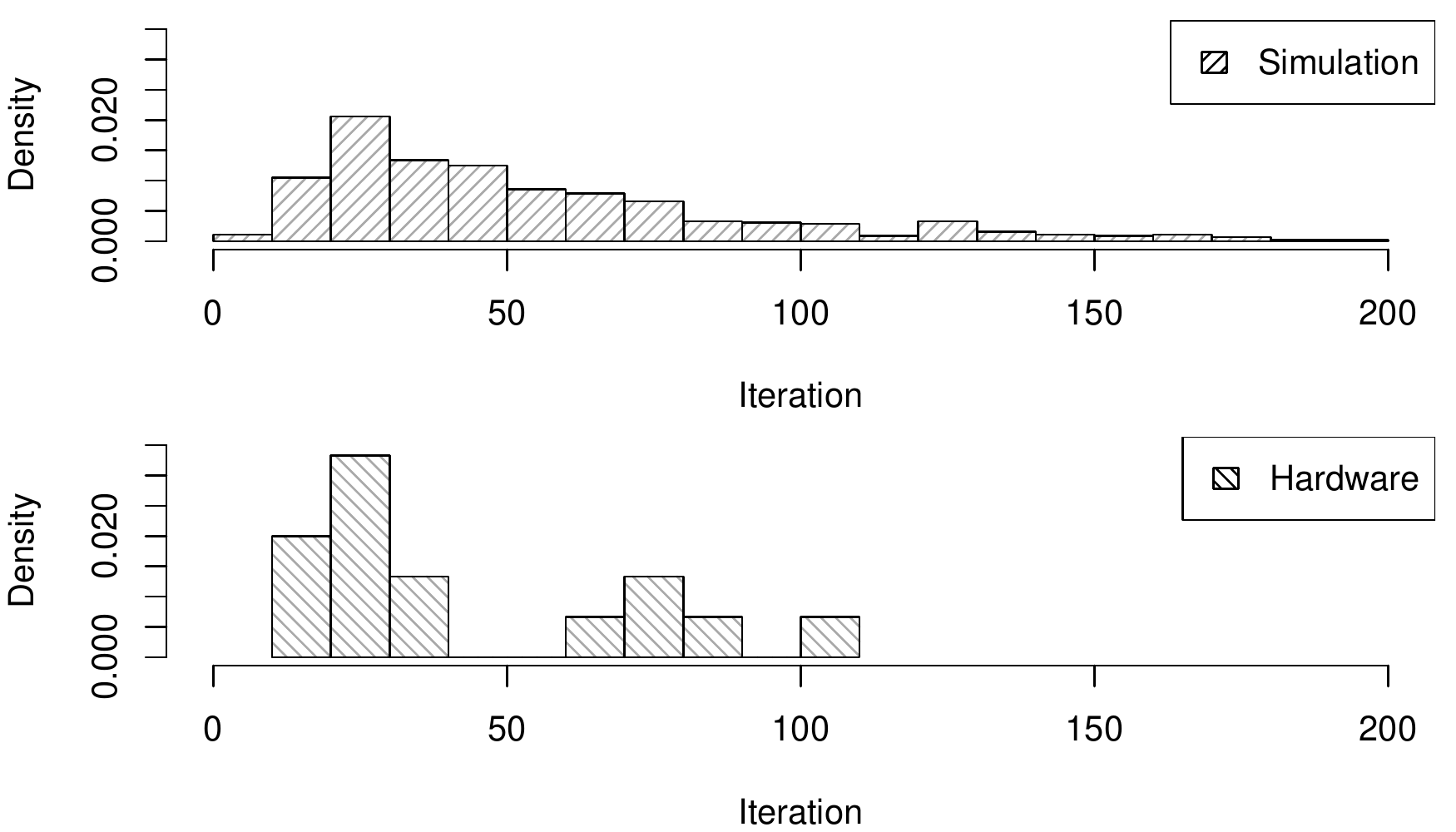}
  \caption{The histogram of the number of steps before the first appearance of a zero tabu matrix starting from a zero matrix at the time origin.}\label{fig1}
\end{figure}

In the simulation experiment we used 10000 repeated runs of the algorithm. Among these, in 457 trajectories there was at least one collision registered in which the matrix $\boldsymbol S$ was identically zero, and the histogram is built for the trajectory length (in terms of the number of iterations $i$) until the first zero (after a zero appeared, iterations were cancelled). The corresponding empirical distribution is depicted on Figure~\ref{fig1} (top).

In the experiment on a hardware, we used 250 repeated runs of length 200 each, at Quantum hardware which was D-Wave computer (Advantage system 4.1, 5000+ cubits, Pegasus topology). Among these, 15 zeroes were obtained, which gives approximately the same rate as in simulation (6\% compared to 4.57\%, respectively). These results required approximately 4.383 minutes of computing time at QPU. The corresponding empirical distribution is depicted on Figure~\ref{fig1} (bottom).

The promising numerical results inspire us to suggest the following \textit{embarrassingly parallel} modification of the Algorithm~\ref{Schema}.  We note the fact that the applications are called embarrassingly parallel when a computational experiment can be decomposed into a huge number independent runs. This specifically fits the distributed computing systems such as being based on BOINC software~\citep{And19}. We follow the approach suggested in~\citep{glynn_topics_1994}, namely, construction of the stochastically equivalent process consisting of a ``stitched'' together independent regeneration periods. Consider we start a huge number of independent copies of TEHQO with the same required parameters. Then we run each trajectory of such a simulation independently until the zeroing of tabu matrix happens. The trajectories in which this zeroing doesn't happen are rejected. Such a simulation is known as time-parallel~\citep{hutchison_tradeoff_2013}. After the simulation completion, necessary statistics are gathered using e.g. regenerative estimation~\citep{glynn_topics_1994}. This also allows to relax Assumption 2 on the TEHQO trajectory stated in Section~\ref{sec3}. However, we leave a detailed analysis of this possibility for future research.

\section{Conclusion and Discussion}
In this paper we presented an approach to the formal proof of convergence of the TEHQO scheme. The scheme generalizes already existing approaches that contain a tabu mastrix structure over an Ising model. The role of the matrix for the proof of asymptotic convergence is clarified in terms of regeneration of the Markov process describing the computation. Empirically, the existence of the regeneration phenomenon is verified in a specific case. However, the speed of convergence and ways to improve the efficiency of the algorithm are to be studied separately. Among the possible ways to continue this research one could consider the tabu matrix parametrization so as to balance the depth of dependency vs. the speed of convergence of the optimization algorithm. Moreover, it might be interesting to perform a comparison with other existing hybrid QC schemes.

\bmhead{Acknowledgments}

The publication has been prepared with the support of Russian Science Foundation according to the research project No.{21-71-10135} \url{https://rscf.ru/en/project/21-71-10135/}. AR would like to acknowledge the support of the research visit which facilitated this research by the Italian CNR STM program.\\
This work was supported by Q@TN, the joint lab between University of Trento, FBK-Fondazione Bruno Kessler, INFN-National Institute for Nuclear Physics and CNR-National Research Council.

\bmhead{Data Availability Statement}
The datasets generated during and/or ana\-lysed during the current study are available from the corresponding author on reasonable request.




\bibliography{qip2022}

\end{document}